\documentclass[aps,prd,onecolumn,eqsecnum,amsmath,nofootinbib,preprintnumbers,10pt]{revtex4}%
\usepackage[dvips]{color,graphicx}
\usepackage{amsfonts,amssymb,theorem,mathrsfs,times}
\usepackage{bm}
\newcommand{\D}{{\rm d}}

{\theorembodyfont{\upshape}
}
{\theorembodyfont{\upshape}
}
{\theorembodyfont{\upshape}
}
{\theorembodyfont{\upshape}
}
{\theorembodyfont{\upshape}
}
{\theorembodyfont{\upshape}
}

\newcommand{\dalm}{\kern1pt\vbox{\hrule height 0.9pt\hbox{\vrule width
0.9pt\hskip 2.5pt\vbox{\vskip 5.5pt}\hskip 3pt\vrule width 0.3pt}\hrule height
0.3pt}\kern1pt}

\begin{document}\large
\preprint{\hfill {\small {ICTS-USTC-15-11}}}
\title{Counterterms in Massive Gravity Theory
}

%

\author{Li-Ming Cao$^{a,b}$, Yuxuan Peng$^{a}$}

\email{caolm@ustc.edu.cn , yxpeng@mail.ustc.edu.cn}


\affiliation{$^a$
Interdisciplinary Center for Theoretical Study\\
University of Science and Technology of China, Hefei, Anhui 230026, China}

\affiliation{$^b$ State Key Laboratory of Theoretical Physics,
Institute of Theoretical Physics, Chinese Academy of Sciences,
P.O. Box 2735, Beijing 100190, China}
%

\date{\today}

\begin{abstract}
We derived local boundary counterterms in massive gravity theory with a negative cosmological constant in four dimensions. With these counterterms at hand we analyzed the properties of the boundary field theory in the context of AdS/CFT duality by calculating the boundary stress energy tensor.
The calculation shows that the boundary stress energy tensor is conserved, and momentum dissipation might occur on the level of linear response only.
We also calculated the thermodynamic quantities and the boundary stress energy tensor for a specific type of solutions. The thermodynamic potentials agree with the results of literature up to some constants which can be removed by adding finite counterterms.
\end{abstract}


\maketitle

\section{introduction}
AdS/CFT correspondence plays an essential role in modern theoretical physics--- it relates a gravity theory in the bulk of an asymptotically AdS spacetime to a conformal field theory without gravity living on the boundary of this spacetime \cite{Maldacena:1997re, Gubser:1998bc, Witten:1998qj, Aharony:1999ti}, and it develops some new methods to study  strongly coupled gauge theories by using gravity theory which is usually weakly coupled. This correspondence is usually referred to as a holography duality, with the two generating functionals satisfying
$Z_{\mathrm{bulk}}(\phi_{0,i}) = Z_{\mathrm{CFT}}(\phi_{0,i})\,, $
where $\phi_{0,i}$ represents the boundary values of the bulk fields propagating in the AdS spacetime, and on the CFT side $\phi_{0,i}$s are the sources coupled to various CFT operators. This equivalence of generating functionals leads to an equivalence of the actions on both sides. When the gravity theory is in the classical limit, the expectation value of the operator $\cal O$ coupled to the source $\phi_{(0)}$ can be computed by the variation of the on-shell action in the bulk, i.e.,
${\langle \cal O} \rangle = \delta S_{\mathrm{bulk}}[\phi_{(0)}]/\delta \phi_{(0)}\,,$
and one example of this is the expectation value of the stress energy tensor
\begin{align}
{\langle T^{\mu\nu}} \rangle = \frac{\delta S_{\mathrm{bulk}}[g_{(0)\mu\nu}]}{\delta g_{(0)\mu\nu}}\,,
\end{align}
with $g_{(0)\mu\nu}$ being the boundary metric tensor. Note that the on-shell action $S$ here has been renormalized. In asymptotically AdS space, without renormalization, the on-shell action is generically divergent as the reader can find in the next section. Actually the bulk theory has one more dimension than the boundary theory and this dimension is just the radial direction of the asymptotically AdS spacetime, representing the energy scale of the boundary theory. The region near the AdS boundary of the bulk spacetime is dual to the ultraviolet region of the boundary field theory. The divergences of the gravity action result from integration over this near-boundary region and are dual to the ultraviolet divergences in the boundary field theory.

In field theory people introduce counterterms in the action to cancel the ultraviolet divergences, and that is renormalization. In the gravity side, one can do similar things by introducing local counterterms on the AdS boundary. In the case of vacuum, the authors of \cite{Balasubramanian:1999re} constructed the local counterterms which are integrals of invariant quantities on the AdS boundary, such as a constant and combinations of Riemann and Ricci tensors. The counterterms are independent of which solution of the Einstein equations is being discussed. With the counterterms added to the gravity action the authors of \cite{Balasubramanian:1999re} derived finite actions and finite boundary stress energy tensors on the field theory side. And the Casimir energy and conformal anomalies of the boundary theory were also computed. The counterterms was also investigated in detail in the paper \cite{Emparan:1999pm}.
The paper \cite{Kraus:1999di} gave a systematic 
method to derive the counterterms from the point of view of boundary stress energy tensor. From viewpoint of the action, the method of constructing the counterterms was analyzed in \cite{Henningson:1998gx, deHaro:2000xn, Bianchi:2001kw, Skenderis:2002wp} and many other papers written by the authors. The authors analyzed the asymptotic behavior of the gravity action and extracted its divergent part, and their derivation included the counterterms for spacetimes containing matter fields.

All of the above are stories in Einstein gravity with a negative cosmological constant. Recently a modified gravity theory--- massive gravity theory has been discussed in many papers (see for instance, \cite{Vegh:2013sk}) in the context of holography to realize momentum dissipation in the field theory on the boundary. However, solution-independent local counterterms for this theory is still unknown. The main task of this paper is to address this problem. Before telling the main story we would like to introduce massive gravity in brief.

From the viewpoint of linear gravity theory, the graviton in Einstein gravity has no mass which can be read out from the well known massless Fierz-Pauli equations.  Fierz and Pauli have introduced a mass for the graviton long time ago \cite{Fierz:1939ix}. However, this kind of massive gravity suffers from the vDVZ discontinuity to Einstein gravity when the mass of graviton approaches zero\cite{Zakharov:1970cc} and from the Boulware-Deser ghost\cite{Boulware:1973my}. The first problem can be solved by the so-called Vainshtein mechanism in some sense\cite{Vainshtein:1972sx}. However, generally,
the Boulware-Deser ghost popularly exits in this kind of theories. Recently, the massive gravity was generalized to a non-linear theory, the so-called dRGT theory, by de Rham, Gabadadze, and Tolley in the work \cite{deRham:2010ik, deRham:2010kj, Hinterbichler:2011tt}. This nonlinear theory was proven to be ghost-free by Hassan and Rosen\cite{Hassan:2011hr,Hassan:2011tf}. For more details the readers can refer to the nice reviews \cite{Hinterbichler:2011tt, deRham:2014zqa}. We just stress one more point that the diffeomorphism invariance, i.e. the coordinate covariance is broken due to the massive terms. In the Lagrangian there is a non-dynamical ``reference metric" $f_{\mu\nu}$ contracted with the spacetime 
metric $g_{\mu\nu}$. Since $f_{\mu\nu}$ is a fixed tensor in any specific case and can not be constructed by $g_{\mu\nu}$, the diffeomorphism is indeed broken. This issue in the context of perturbations was also discussed in the paper \cite{Davison:2013jba}. Now that the bulk diffeomorphism of the  gravity theory has been broken, according to the AdS/CFT correspondence, the dual boundary field theory loses the global translational symmetry. So a finite DC conductivity  can be realized by a black hole solution with finite temperature and charge in this massive gravity\cite{Vegh:2013sk, Blake:2013bqa}.

With solution-independent counterterms at hand, we can look into the renormalized boundary stress energy tensor and see if there is any signs of momentum dissipation. This helps us understand the mechanism of symmetry breaking more thoroughly. In this paper, we derived the counterterms and computed the renormalized boundary stress energy tensor and find that momentum conservation is not violated at least on the background level. Our derivation included the divergent part of the action, while the issue of finite counterterms had been discussed in the paper \cite{Amoretti:2014zha}. 

The thermodynamics and phase transition of the massive gravity model were studied in the papers \cite{Cai:2014znn} and the case in the extended phase space was studied in \cite{Xu:2015rfa}. With some special assumptions, the authors of the paper \cite{Hu:2015xva} derived the Misner-Sharp mass and they also  studied the thermodynamics. In the paper \cite{Cai:2014znn} the theromdynamic quantities were derived by using the Hamiltonian approach. In this paper with the counterterm action we derive the thermodynamic quantities by computing the renormalized Euclideanized on-shell action. We find that the thermodynamic potentials are the same, up to some constants, as in \cite{Cai:2014znn}. The constants result from the massive terms and can be removed by adding finite counterterms.

This paper is organized as follows. In the next section we introduce briefly the method of constructing counterterms in Einstein gravity. The counterterms for massive gravity are derived in Section \ref{main} and the conservation of the boundary stress energy tensor is analyzed. In section \ref{app of counter} we apply the counterterms to a specific solution and derive the thermodynamic quantities to compare with the results in the literature.  At the end we give some conclusions. The details of computation are given in the appendices.

\section{Counterterms in Einstein gravity}

The method of constructing the local counterterms in  Einstein gravity in $(d+1)$ dimensions is briefly introduced in this section according to the renowned papers \cite{Henningson:1998gx, deHaro:2000xn, Bianchi:2001kw, Skenderis:2002wp}, etc.. The procedure consists of expanding the on-shell action with respect to powers in a small parameter $\epsilon$ near the AdS boundary and re-expressing the divergent part of the on-shell action in terms of the invariants on the boundary by solving the Einstein equations.

Asymptotically AdS spacetimes or more general asymptotically locally AdS spacetimes in $(d+1)$-dimensions are solutions of the action
\begin{equation}
S_{\mathrm{EG}}={1 \over 2\kappa^2}\int_{M}\mathrm{d}
^{d+1}x\,
\sqrt{-\mathcal{G}}\, (R[\mathcal{G}] - 2 \Lambda) -\frac{1}{\kappa^2} \int_{\partial M} \mathrm{d}^d x\, \sqrt{-\gamma}\, K\,,
\end{equation}
where $K$ is the trace of the second fundamental form and
$\gamma$ is the induced metric on the boundary, $\partial M$, of the spacetime region $M$. The
boundary term is necessary in order to get an action which
only depends on first derivatives of the metric \cite{Gibbons:1976ue},
and it guarantees that the variational
problem with Dirichlet boundary conditions is well-defined. We further set $\Lambda = -d(d-1)/(2\ell^2)$ and set $\ell=1$ in most of this paper. We will recover $\ell$ later.
The metric in the neighborhood of the boundary of an asymptotically AdS spacetime takes the form
\begin{equation}
\label{coord in u}
\mathcal{G}_{\mu\nu}\mathrm{d}x^\mu \mathrm{d}x^\nu =\frac{1}{u^2} \Big(\mathrm{d} u^2 + g_{ij}(x,u) \mathrm{d}x^i \mathrm{d}x^j\Big)\, ,\quad i, j=0\,, 2\, , ...\, , d+1\, ,
\end{equation}
where
\begin{equation}
\label{expan in u}
g_{ij}(x,u)=g_{(0)ij} + g_{(1)ij}u + g_{(2)ij}u^2 + \cdots
\end{equation}
The coordinate $x^1=u$ is the radial metric and $g$ is a ``metric" in $d$ dimensions. The AdS boundary is located at $u=0$. So far a specific $(3+1)$- decomposition is given and the induced metric on the hypersurface of constant $u$, denoted by $\gamma_{\mu\nu}$, is given by
\begin{align}
\gamma_{\mu\nu}\mathrm{d}x^{\mu}\mathrm{d}x^{\nu}=\gamma_{ij}\mathrm{d}x^{i}\mathrm{d}x^{j} = {1 \over u^2}g_{ij}\mathrm{d}x^{i}\mathrm{d}x^{j}\,,
\end{align}
and the outward-pointing normal vector can be expressed as
\begin{align}
n_{\mu}\mathrm{d}x^{\mu} = - \frac{\D u}{u}\,.
\end{align}
By this seting,  the extrinsic curvature $K_{\mu\nu}$ is defined as half the Lie derivative
\begin{align}
K_{\mu\nu}\mathrm{d}x^{\mu}\mathrm{d}x^{\nu}=K_{ij}\mathrm{d}x^{i}\mathrm{d}x^{j} = -\frac{1}{2} (\mathcal{L}_n \gamma_{ij})\mathrm{d}x^{i}\mathrm{d}x^{j}\,.
\end{align}
This coordinate system is actually a Gaussian normal coordinate system with ${\partial}/{\partial u}$ generating geodesics with respect to the ``metric" $u^2\mathcal{G}_{\mu\nu}$.
Throughout this paper the lattin indices $i,j$ and $k$, etc. denote the coordinates on the hypersurface of constant radius $u$ and the Greek indices $\mu, \nu,$ and $\alpha$, etc. denote all the coordinates. Explicit computation shows that in pure gravity all coefficients multiplying odd powers of $u$ vanish up to the order $u^d$. So one can use the coordinate $\rho = u^2$ instead and the metric can be rewritten as
\begin{eqnarray}
 \label{coord}
&&\mathcal{G}_{\mu \nu} \mathrm{d}x^\mu \mathrm{ d}x^\nu = {d\rho^2 \over 4 \rho^2} +
{1 \over \rho} g_{ij}(x,\rho) \mathrm{d}x^i \mathrm{d}x^j\, ,\nonumber\\
&&g(x,\rho)=g_{(0)} + \rho^2 g_{(2)} + \cdots + \rho^{d/2} g_{(d)} + h_{(d)} \rho^{d/2} \log \rho + ... \, .
\end{eqnarray}
The expansion above contains a logarithmic term and it is related to the conformal anomaly in the boundary field theory and it only appears in the cases of even $d$. Using the vacuum Einstein equations\footnote{Our convention for $R_{\mu\nu}$ is different from that in \cite{Henningson:1998gx} by a minus sign.}
\begin{equation}
R_{\mu\nu} - \frac{1}{2}R \mathcal{G}_{\mu\nu} + \Lambda \mathcal{G}_{\mu\nu} = 0
\end{equation}
one can express the action as:
\begin{eqnarray}
 \label{action}
S_{\mathrm{EG}}&=&{1 \over 2\kappa^2}\left[\int_{M}\mathrm{d}^{d+1}x\,
\sqrt{-\mathcal{G}}\, (R[\mathcal{G}] - 2 \Lambda)
- \int_{\partial M} \mathrm{d}^d x\, \sqrt{-\gamma}\, 2 K\right] \nonumber\\
&=&{1 \over 2\kappa^2} \int \mathrm{d}^d x \Bigg{\{}
-\int_\epsilon \mathrm{d}\rho\,{d\over\rho^{d/2+1}}\,\sqrt{- g(x,\rho )}
\nonumber\\
&&+ \left.{1\over \rho^{d/2}}
\left(2 d \sqrt{- g(x,\rho )}
-4 \rho\partial_\rho \sqrt{- g (x,\rho )}\right)\right|_{\rho=\epsilon}\Bigg{\}}\, .
\end{eqnarray}
This action is regulated by restricting
the bulk integral to the region $\rho\geq\epsilon$
and evaluating the boundary term at $\rho=\epsilon$. It can be expanded in powers in $\epsilon$ plus a logarithmic term:
\begin{eqnarray}
 \label{regaction1}
S_{\mathrm{EG,reg}} &=&{1 \over 2\kappa^2} \int \mathrm{d}^d x \sqrt{- g_{(0)}} \left[
\epsilon^{-d/2} a_{(0)} + \epsilon^{-d/2+1} a_{(2)} + \ldots
+ \epsilon^{-1} a_{(d - 2)} \right. \nonumber\\
& & \qquad \left.  -\log \epsilon\, a_{(d)} \right] + \mathcal{O}(\epsilon^0)\,.
\end{eqnarray}
The counterterm action is equal to the divergent part with a minus sign added:
\begin{eqnarray}
S_{\mathrm{ct,reg}} &=& -{1 \over 2\kappa^2} \int \mathrm{d}^d x \sqrt{- g_{(0)}} \left[
\epsilon^{-d/2} a_{(0)} + \epsilon^{-d/2+1} a_{(2)} + \ldots
+ \epsilon^{-1} a_{(d - 2)}\right. \nonumber\\
& & \qquad\quad \left.  - \log \epsilon\, a_{(d)} \right] \,.
\end{eqnarray}
The renormalized action will be
\begin{equation}
S_{\mathrm{ren}} = \lim_{\epsilon\rightarrow 0}\big( S_{\mathrm{EG,reg}} + S_{\mathrm{ct,reg}}\big)\,.
\end{equation}
The coefficients $a_{(i)}, i=0,2,...$ can be expressed by the coefficients $g_{(i)ij}$. In order to re-express the counterterms in terms of invariant quantities on the boundary, one can consider the Einstein equations in the decomposed form:
\begin{eqnarray}
\label{eom in gr}
&&\rho \,[2 g^{\prime\prime} - 2 g^\prime g^{-1} g^\prime + \mathrm{Tr}\,
(g^{-1} g^\prime)\, g^\prime] - {\rm Ric} (g) - (d - 2)\,
g^\prime - \mathrm{Tr} \,(g^{-1} g^\prime)\, g = 0\, , \nonumber\\
&&\mathrm{Tr} \,(g^{-1} g^{\prime\prime}) - \frac{1}{2} \mathrm{Tr} \,(g^{-1} g^\prime
g^{-1}
g^\prime) = 0  \, ,\nonumber\\
&&\hat{D}_i\, \mathrm{Tr} \,(g^{-1} g^\prime) - \hat{D}^j g_{ij}^\prime  = 0\,,
\end{eqnarray}
where differentiation with respect to $\rho$ is denoted with a prime, and
$\hat{D}_i$ is the covariant derivative constructed from the metric
$g$, and ${\rm Ric} (g)$ is the Ricci tensor of $g$. Solving these equations order by order gives the relations between the coefficients $a_{(i)}$ and the geometric quantities on the boundary, up to finite terms. Then one can write down the counterterm action. The final results for generic dimensions is given by \cite{deHaro:2000xn}
\begin{eqnarray}
\label{ct}
S_{\mathrm{ct}}&=&-{1 \over 2 \kappa^2} \int_{\rho=\epsilon}
\sqrt{-\gamma}\left[2(d-1) + {1 \over d-2} R + {1 \over (d-4) (d-2)^2}
(R_{ij} R^{ij} - {d \over 4 (d-1)} R^2)\right. \nonumber\\
& & \qquad \quad \left.  - \log \epsilon\, a_{(d)} + ...\right]\,,
\end{eqnarray}
with $a_{(d)}$ determined by the conformal anomaly. This expression is enough up to $d=6$.

\section{Counterterms for massive gravity}
\label{main}
We work in four dimensions, with the massive gravity model studied in \cite{Vegh:2013sk, Davison:2013jba, Blake:2013bqa} and \cite{Cai:2014znn}. The action of the theory has a form
\begin{equation}
S_{\mathrm{mg}}=\frac{1}{2 \kappa^2}\int_M d^4 x \sqrt{-\mathcal{G}}\left[R + 6  + m^2\left( c_1 {\cal U}_1 + c_2 {\cal U}_2\right) \right]- \frac{1}{\kappa^2}\int_{\partial M} d^3 x \sqrt{-\gamma} K\, .
\end{equation}
We follow the conventions in \cite{Cai:2014znn}, and we only consider the pure gravity case for simplicity. In the action $m$, $c_1$ and $c_2$ are constants. There are four massive terms $c_i {\cal U}_i, i=1, 2, 3, 4$ in \cite{Cai:2014znn}, but in the case of this paper only the first two coefficients $c_1$ and $c_2$ survive. The functions ${\cal U}_1$ and ${\cal U}_2$ are constructed by the eigenvalues of the $(1,1)-$type tensor ${{\cal K}^\mu}_\nu$
\begin{eqnarray}
\mathcal{ U}_1&=& {\cal K}^{\mu}{}_{\mu}\, , \nonumber \\
\mathcal{ U}_2&= &({\cal K}^{\mu}{}_{\mu})({\cal K}^{\nu}{}_{\nu})-\mathcal{K}^{\mu}{}_{\nu}\mathcal{K}^{\nu}{}_{\mu}\, ,
\end{eqnarray}
where the tensor ${\cal K}^\mu{}_\nu$ is defined by ${\cal K}^\mu{}_\alpha{\cal K}^\alpha{}_\nu \equiv ({\cal K}^2)^\mu{}_\nu \equiv \mathcal{G}^{\mu\alpha}f_{\alpha\nu} $. The tensor $f_{\mu\nu}$  is called the ``reference metric", a fixed symmetric tensor breaking the diffeomorphism invariance of the theory. We follow the method in \cite{deHaro:2000xn}. In the case of massive gravity, the space-time is not asymptotically AdS in
the usual sense \cite{Blake:2013bqa, Cai:2014znn}. Instead of $\rho$, the coordinate $u$ will be used. The coordinate system is the same as (\ref{coord in u}), i.e.,
\begin{align}
\mathcal{G}_{\mu\nu}\D x^{\mu}\D x^{\nu} = \frac{1}{ u^2}\big(d u^2 + g_{ij}(x,u) \mathrm{d}x^i \mathrm{d}x^j\big)  \, , \quad i=0\, , 2\, , 3\, ,
\end{align}
and the expansion is exactly the same as (\ref{expan in u}). Here we include the odd powers in $u$ in the expansions, and the reason can be found in the behavior of the equations of motion later. To solve the equations of motion consistently we do not need include logarithmic terms here since $d$ is odd. We have to mention that not like \cite{Vegh:2013sk, Davison:2013jba, Blake:2013bqa} and \cite{ Cai:2014znn} where the form of $f_{\mu\nu}$ is exactly given, in this paper, we only impose some conditions on the massive terms.

Here come the conditions. We assume that the spacetime $(M,\mathcal{G})$ can be foliated by a family of 2-dimensional spacelike surfaces one of which we are considering is denoted by $\Sigma$, a submanifold of a constant $u$ hypersurface, and this means that the metric of the spacetime can always be decomposed  into the form
\begin{equation}
\mathcal{G}_{\mu\nu}\mathrm{d}x^{\mu}\mathrm{d}x^{\nu} =\big(- t_\mu t_\nu+ n_\mu n_\nu  + \sigma_{\mu\nu}\big)\mathrm{d}x^{\mu}\mathrm{d}x^{\nu}  \,,
\end{equation}
in which $\sigma_{\mu\nu}\mathrm{d}x^{\mu}\mathrm{d}x^{\nu}$ is the induced metric of the spacelike surface $\Sigma$. The covector $t_\mu \mathrm{d}x^{\mu}$ is timelike and normal to $\Sigma$ and normal to $n_\mu \mathrm{d}x^{\mu}$, and $n_\mu \mathrm{d}x^{\mu}$ is the unit normal which has been introduced in the previous section. This means that $\sigma_{\mu\nu} n^\nu=0$ and $\sigma_{\mu\nu} t^\nu=0$. Actually we have decomposed the constant $u$ hypersurface into $t^\mu$ and $\Sigma$ here. We further impose the conditions that ${{\cal K}^\mu}_\nu$ and $({\cal K}^2)^\mu{}_\nu$ satisfy
\begin{equation}
\label{degenerating K}
{{\cal K}^\mu}_\nu = \frac{1}{2}({\cal K}^{\alpha}{}_{\alpha}){\sigma^\mu}_\nu\,,\qquad {f^\mu}_\nu = ({\cal K}^2)^\mu{}_\nu = \frac{1}{2} \big(({\cal K}^2)^\alpha{}_\alpha\big) {\sigma^\mu}_\nu\,.
\end{equation}
In the case that $t^\mu \equiv \mathcal{G}^{\mu\nu}t_\nu = (\partial/\partial t)^{\mu}$, the above conditions mean that only the diffeomorphism invariance on $\Sigma$ is broken.  The boundary dual of such a system has lost the translational symmetry on $\Sigma$ \cite{Davison:2013jba, Blake:2013bqa}.

From the above we have that the components $\mathcal{K}_{u\rho}$ are vanishing. Therefore we can write  ${\cal K}_{\mu\nu}$ as ${\cal K}_{ij}$, and
\begin{equation}\label{trace squared}
{\cal U}_1^2 = ({\cal K}^{i}{}_{i})^2 = 2({\cal K}^2)^{i}{}_{i}\,.
\end{equation}
A simple assumption which we make is that the expansions of the massive terms near the boundary are
\begin{eqnarray}\label{expan of ma terms}
{\cal K}_{ij}&\equiv &\frac{1}{u}\xi_{ij} = \frac{1}{u}\left[\xi_{(0)ij} + u\xi_{(1)ij} + \cdots\right]\,, \\
{\cal U}_1 &\equiv& u\mu = u\left[ \mu_{(0)} + \mu_{(1)}u + \cdots\right]=u{\xi^i}_i\,, \\
{\cal U}_2 &\equiv& u^2\lambda = u^2\left[ \lambda_{(0)} + \lambda_{(1)}u + \cdots\right]\,.
\end{eqnarray}
The equations  of motion turn out to be
\begin{equation}\label{eom}
R_{\mu\nu}-\frac{1}{2}R \mathcal{G}_{\mu\nu}-3 \mathcal{G}_{\mu\nu}=\frac{1}{2}m^2\left[c_1\left({\cal U}_1 \mathcal{G}_{\mu\nu}-{\cal K}_{\mu\nu}\right)+c_2\left({\cal U}_2 \mathcal{G}_{\mu\nu}-2{\cal U}_1{\cal K}_{\mu\nu}+2({\cal K}^2)_{\mu\nu}\right)\right]\, .
\end{equation}
Note that the equations of motion can only be derived after specifying the rank of the matrix $f_{\mu\nu}$. Here $f_{\mu\nu}$ is degenerate. However, the equations (\ref{eom}) are written in the same form as in the case in which $f_{\mu\nu}$ has full rank. This is permitted under the conditions (\ref{degenerating K}).
Similar to Eq.(\ref{eom in gr}), the equations of motion (\ref{eom}) imply that
\begin{eqnarray}
&&-\frac{1}{4} \mathrm{Tr}\left(g^{-1}g'\right)g'_{ij} - \frac{1}{2} g''_{ij} + \frac{1}{2} \left(g'g^{-1}g'\right)_{ij} + \frac{1}{2u}\mathrm{Tr}\left(g^{-1}g'\right)g_{ij} + \frac{1}{u} g'_{ij} + R_{ij}(g) \nonumber\\
&&= m^2 \left[ c_1\left(-\frac{{\cal U}_1}{4u^2}g_{ij} - \frac{{\cal K}_{ij}}{2}\right) + c_2\left(- {\cal U}_1 {\cal K}_{ij} + {\cal K}^2_{ij}\right) \right]\,, \label{evolution}
\end{eqnarray}
\begin{eqnarray}\label{metric divergence eq}
\hat{D}_i\, \mathrm{Tr} \,(g^{-1} g^\prime) - \hat{D}^j g_{ij}^\prime  & = & 0\,.
\end{eqnarray}
\begin{eqnarray}
-\frac{u^2}{2}\mathrm{Tr}\left(g^{-1}g''\right) + \frac{u^2}{4}\mathrm{Tr}\left(g^{-1}g'g^{-1}g'\right) + \frac{u}{2}\mathrm{Tr}\left(g^{-1}g'\right) = -\frac{1}{4}m^2c_1{\cal U}_1\, .\label{trace}
\end{eqnarray}
The prime $``\, '\, "$ represents the derivative with respect to $u$, and $R_{ij}(g)$ is the Ricci tensor constructed by $g_{ij}$. Now we can see that odd powers in $u$ in the tensor $g_{ij}$ can render the equations of motion consistent.

Similar to the case of Einstein gravity, the on-shell action can be expanded as
\begin{eqnarray}
S_{\mathrm{mg,reg}}& =&\, \frac{1}{2\kappa^2} \int \mathrm{d}^3 x \int_{\epsilon} \mathrm{d}u \sqrt{-g}\left( -\frac{6}{\ell^2} - \frac{m^2 c_1 {\cal U}_1}{2}\right) - \frac{1}{\kappa^2} \int \mathrm{d}^3 x \sqrt{-\gamma} K \nonumber \\
&=& \frac{1}{2\kappa^2} \int \mathrm{d}^3 x \left[ a_{(0)} \epsilon^{-3} + a_{(1)} \epsilon^{-2} + a_{(2)} \epsilon^{-1} + a_{(3)} \ln \epsilon + \mathcal{O}(\epsilon^0)\right] \,,
\end{eqnarray}
where the small parameter $\epsilon$ denotes small $u$ and the coefficients $a_{(i)}, i=0,1,2,3$ are given in the appendix \ref{appcoeff}. To obtain the expressions for them we can solve eq.(\ref{evolution}) and eq.(\ref{trace}) order by order. The solutions to the order high enough have been put into appendix \ref{appeom}.

What is worth mentioning is that the coefficient $a_{(3)}$ of the logarithmic divergence vanishes, proven in the appendix \ref{appcoeff}. Therefore we just need to cancel the power law divergences in the action. Finally, the counterterm with $\ell$ recovered is
\begin{align}\label{countertermaction}
S_{\mathrm{ct}}&= -\frac{1}{\kappa^2} \int \mathrm{d}^3 x \sqrt{-\gamma}\left[\frac{2}{\ell} + \frac{1}{2} \ell R[\gamma] + \frac{1}{4}m^2 c_1 \ell \, {\cal U}_1 + \left(\frac{1}{2} m^2 c_2 \ell - \frac{1}{32}m^4 c_1^2 \ell^3\right) {\cal U}_2  \right]
\end{align}
The terms $2/\ell$ and $\ell R[\gamma]/2$, where $R[\gamma]$ is the scalar curvature of the induced metric $\gamma_{\mu\nu}$, are familiar: they already exist in the case of Einstein gravity theory. The last three terms are new in the massive gravity. We have to stress that this counterterm is valid only under the conditions we imposed before. The functions ${\cal U}_1$ and ${\cal U}_2$ are indeed functions on the 3-boundary, since ${\cal K}_{\mu\nu}$ is actually ${\cal K}_{ij}$ under those conditions. So this counterterm action is constructed by boundary invariant quantities only.

 Since the general form of the renormalized action is at hand, we can compute the variation with respect to the boundary metric $\gamma_{\mu\nu}$ of the on-shell action. The result is the renormalized Brown-York stress-energy tensor~\cite{Brown:1992br} for this theory:
\begin{equation}\label{B-Y def}
T^{\mu\nu} = \frac{2}{\sqrt{-\gamma}}\frac{\delta (S+S_{\mathrm{ct}})}{
\delta \gamma_{\mu\nu}} \, ,
\end{equation}
Since the stress energy  tensor is a tensor on the boundary,  $T_{ij}$ are the only nontrivial components of the tensor and  the tensor consists of two parts, i.e.,
\begin{equation}
T_{ij}= T^{\mathrm{EG}}_{ij} + T^{\mathrm{mg}}_{ij} \,.
\end{equation}
The first part is the contribution from Einstein gravity
\begin{equation}
T^{\mathrm{EG}}_{ij} =\frac{1}{\kappa^2}\Big[ K_{ij} - \gamma_{ij}K - \gamma_{ij}\frac{2}{\ell} + \ell\,\Big(R[\gamma]_{ij} - \frac{1}{2} \,R[\gamma]\gamma_{ij} \Big)\Big]\,,
\end{equation}
and the second part comes from the massive terms
\begin{eqnarray}
T^{\mathrm{mg}}_{ij}& =&  -\frac{1}{\kappa^2} \Big[\,\frac{1}{4}m^2 c_1 \ell \left({\cal U}_1 \gamma_{ij} - {\cal K}_{ij}\right) \nonumber\\
& & + \left(\frac{1}{2} m^2 c_2 \ell - \frac{1}{32}m^4 c_1^2 \ell^3 \right) \left( {\cal U}_2 \gamma
_{ij} - 2 {\cal U}_1 {\cal K}_{ij} + 2 {\cal K}_{ik}{{\cal K}^k}_j \right) \Big]\,.
\end{eqnarray}
One can check that the divergences in the expansion with respect to small $u$ of the tensor $T_{ij}$ are canceled out. So is the constant term, under the assumptions (\ref{degenerating K}).  Therefore the leading term in the expansion is of order $u$.

The 3-dimensional divergence of $T_{ij}$ is
\begin{eqnarray}
\label{divergence of energy tensor}
D^i T_{ij} &=& D^i T^\mathrm{EG}_{ij} + D^i T^\mathrm{mg}_{ij} \nonumber\\
&=& \frac{m^2 \ell}{\kappa^2} \Big[\, \frac{1}{4} c_1 D^i \left( {\cal K}_{ij}
- {\cal U}_1\gamma_{ij}\right)\nonumber\\
& & + \left(c_2 - \frac{1}{16} m^2 c_1^2 \ell^2\right) D^i \left( -\frac{1}{2} {\cal U}_2 \gamma_{ij} + {\cal U}_1 {\cal K}_{ij} - {\cal K}_{ik}{{\cal K}^k}_j\right) \Big] \nonumber\\
&=& 0 \,.
\end{eqnarray}
The operator $D_i$ is the covariant derivative compatible with $\gamma_{ij}$. The divergence is vanishing given
the conservation of the RHS of (\ref{eom}) and that ${\partial}/{\partial u}$ generates geodesics with respect to the ``metric" $u^2\mathcal{G}_{\mu\nu}$.
The vanishing of the divergence of  $T_{ij}$ suggests that we can define some conserved quantities $Q_{\xi}$ associated with some Killing vectors $\xi$
on the boundary
\begin{equation}
Q_{\xi}=\int \D^2 x \sqrt{\sigma}  (\partial/\partial t)^i T_{ij}\xi^j /N_\Sigma \,,
\end{equation}
where $\sqrt{\sigma}$ is the volume element of the $2$-surface with the coordinates $x,y$ and $N_{\Sigma}$ is the lapse function in the ADM decomposition of the boundary spacetime~\cite{Balasubramanian:1999re}. This is a conserved quantity because on any two different slices of $t$ its value remains the same due to (\ref{divergence of energy tensor}).

According to the AdS/CFT duality, the tensor $T_{ij}$ corresponds to the expectation value of the renormalized stress energy tensor $\langle \hat{T}_{ij}\rangle$ of the field theory on the boundary in the following way:
\begin{equation}
\label{Tmn on the boundary}
\langle \hat{T}_{ij}\rangle = \lim_{u\rightarrow 0} \frac{ T_{ij}}{u}\,.
\end{equation}
So the divergence of the stress energy  tensor on the boundary field theory is zero:
\begin{align}
\hat{D}^i \langle \hat{T}_{ij}\rangle =0\,,
\end{align}
with $\hat{D}$ the covariant derivative in the spacetime in which the field theory lives. It seems that momentum conservation is not violated. However, according to \cite{Vegh:2013sk, Davison:2013jba, Blake:2013bqa}, the diffeomorphism invariance in the bulk is broken and so is the translational invariance on the boundary, and there is momentum dissipation in the context of perturbations. So there is indeed momentum dissipation, but not manifest on the background level.

\section{Thermodynamic quantities}\label{app of counter}
In this section we apply the counterterm (\ref{countertermaction}) to the black hole solutions in \cite{Cai:2014znn}. Again we work in 4-dimensions. The action including Maxwell field has a form
\begin{eqnarray}
S &=&\,\frac{1}{2 \kappa^2}\int_M d^4 x \sqrt{-\mathcal{G}}\left[R + \frac{6}{\ell^2} -\frac{1}{4} F^2 + m^2\left( c_1 {\cal U}_1 + c_2 {\cal U}_2\right) \right] \nonumber\\ 
& & - \frac{1}{\kappa^2}\int_{\partial M} d^3 x \sqrt{-\gamma} K \,.
\end{eqnarray}
The metric is given by
\begin{equation}
ds^2 = - f(r) \D t^2 + f^{-1}(r) \D r^2 + r^2 h_{ij} \D x^i \D x^j \, ,
\end{equation}
where $h_{ij}\D x^i\D x^j$ is the line element for a $2$-dimensional Einstein space with constant curvature $2k, k=-1, 0, +1$. The boundary is at $r\rightarrow \infty$. The case of $k=0$ is also studied by \cite{Vegh:2013sk, Davison:2013jba, Blake:2013bqa}. The function $f(r)$ is given by
\begin{equation}
f(r) = k +\frac{r^2}{\ell^2} -\frac{m_0}r+\frac{ \mu^2 r_+^{2}}{4r^{2} } +  \frac{c_1m^2}{2}r + c_2m^2\, ,
\end{equation}
where the constant $m_0$ can be expressed by the outer horizon radius $r_+$ as
\begin{align}
m_0 = k r_+ +\frac{r_+^3}{\ell^2}+\frac{ \mu^2 r_+}{4 } + \frac{c_1m^2 }{2 }r_+^2 + c_2m^2 r_+\,.
\end{align}
In this model the electric field is included, giving the chemical potential $\mu$. However the electric field doesn't produce divergences in the action or in the Brown-York stress-energy tensor. The reference metric is given by
\begin{equation}
\label{reference}
f_{\mu\nu}\D x^{\mu}\D x^{\nu} =  h_{ij}\D x^i \D x^i\, ,
\end{equation}
and the massive terms are
\begin{eqnarray}
{\cal U}_1&=& \frac{2}{r} \, , \nonumber \\
{\cal U}_2&=& \frac{2}{r^2}\, .
\end{eqnarray}
This model satisfies our conditions (\ref{degenerating K}) and (\ref{expan of ma terms}).
We are going to study the effects of the counterterm (\ref{countertermaction}) in this model. It is not hard to find that our proposal (\ref{countertermaction}) evaluated on this solution is
\begin{equation}
S_{\mathrm{ct}} = - \frac{1}{\kappa^2} \int \mathrm{d}^3 x \sqrt{-\gamma}\left[\frac{2}{\ell} + \frac{\ell k}{r^2} + \frac{1}{2} m^2 c_1 \frac{\ell}{r}  -\frac{1}{16} m^4 c_1^2 \frac{\ell^3}{r^2}   + m^2 c_2 \frac{\ell}{r^2} \right]\,.
\end{equation}
Our counterterm has indeed canceled the divergences in the onshell action of this specific solution. The total action $S+S_{\mathrm{ct}}$ is rendered finite. By the way, when $k=0$ this specific form of $S_{\mathrm{ct}}$ is the same as proposed in \cite{Blake:2013bqa}. The grand potential with fixed chemical potential $\mu$ is given by the renormalized Euclidean action:
\begin{align}
\Omega= T S_\mathrm{E} =&\, \frac{V}{2\kappa^2}\left( k r_+ -\frac{r_+^3}{\ell^2}+c_2m^2 r_+ -\frac{1}{4}\mu^2 r_+ \right) \nonumber\\
&\, + \frac{V}{\kappa^2}\left( \frac{m^4 c_1 c_2 \ell^2}{4} - \frac{m^6 c_1^3 \ell^4}{64} + \frac{k m^2 c_1 \ell^2}{4}\right)\,,
\end{align}
where $V$ is the volume of the $2$-surface with the metric $h_{ij}$. This form is the same as in \cite{Cai:2014znn} up to the constant terms without $r_+$. Computation of the temperature, entropy and total electric charge is straightforward
\begin{eqnarray}
T &=& -\frac{f'(r_+)}{4\pi}= \frac{1}{4\pi r_+} \left( k+ 3\frac{r_+^2}{\ell^2} -\frac{1}{4}\mu^2 +c_1m^2r_+ + c_2 m^2 \right)\, , \\
S &=& -\left(\frac{\partial\Omega}{\partial T}\right)_\mu = \frac{2\pi V}{\kappa^2}r_+^2\, ,\\
Q &=& -\left(\frac{\partial\Omega}{\partial \mu}\right)_T = \frac{V \mu r_+}{2 \kappa^2}\, .
\end{eqnarray}
The total energy of the system can be derived by the thermodynamic relation:
\begin{eqnarray}\label{E of thermo}
E &=& \Omega + TS + \mu Q \nonumber\\
&=& \frac{V }{\kappa^2}\left(k r_+ + \frac{r_+^3}{\ell^2} +\frac{ \mu^2 r_+}{4 } + \frac{c_1 m^2 r_+^2}{2} + c_2 m^2 r_+  \right)\nonumber\\
& &+ \frac{V}{\kappa^2}\left( \frac{m^4 c_1 c_2 \ell^2}{4} - \frac{m^6 c_1^3 \ell^4}{64} + \frac{k m^2 c_1 \ell^2}{4} \right) \, .
\end{eqnarray}
This also agrees with \cite{Cai:2014znn} up to the constant terms.
We can also derive the conserved charge associated with $\partial / \partial t$ via the method in \cite{Balasubramanian:1999re} since the system preserves the translational symmetry in the timelike direction.
For the solutions, the components of the renormalized Brown-York stress-energy tensor are computed by (\ref{B-Y def}):
\begin{eqnarray}
\kappa^2 T_{tt} &=&  \frac{\varepsilon }{r} + \mathcal{O}(r^{-2})\,, \nonumber\\
\kappa^2 T_{ij} &=& \frac{\varepsilon}{2 r} \ell^2 h_{ij}   + \mathcal{O}(r^{-2})\,, \quad i,j= 2, 3\, ,
\end{eqnarray}
where $\varepsilon$ is defined as
\begin{equation}
\varepsilon=\frac{1}{\ell}\left(k r_+ +\frac{r_+^3}{\ell^2}+\frac{ \mu^2 r_+}{4 } + \frac{c_1m^2 }{2 }r_+^2 + c_2m^2 r_+\right) + \frac{m^4 c_1 c_2 \ell}{4} - \frac{m^6 c_1^3 \ell^3}{64} + \frac{k m^2 c_1 \ell}{4} \, .
\end{equation}
Here we can find that the divergences and the constant part in the stress energy tensor have been canceled out.
The conserved charge is given by the integration
\begin{align}
E= \int_{r\rightarrow \infty} \mathrm{d}^2 x \sqrt{\sigma} T_{tt} /N_{\Sigma}\,.
\end{align}
This gives exactly the result of eq.(\ref{E of thermo}). 
By the way, the expression of the components show that $\langle \hat{T}_{\mu\nu}\rangle$  is traceless, so the boundary field theory is still conformal even in massive gravity.

In a word, our counterterms give the same thermodynamic relations as in the literature, but there are finite constant terms. These finite terms can be removed by redefining the counterterm action as
\begin{equation}
S_{\mathrm{ct}} = - \frac{1}{\kappa^2}\int d^3 x \sqrt{-\gamma} \sqrt{\frac{4}{\ell^2} + 2R[\gamma] +  m^2 \big(c_1{\cal U}_1 + 2  c_2{\cal U}_2}\big)\,,
\end{equation}
which can give the exact form suggested by \cite{Blake:2013bqa} for the exact solution studied in this paper. This counterterm action is actually (\ref{countertermaction}) plus some higher order finite terms. These terms are somewhat arbitrarily chosen, and whether there is any physical meaning behind is yet to be investigated.

\section{conclusions}
In ghost-free massive gravity theory, the negative cosmological constant will lead to some solutions which have timelike boundaries in the infinity\cite{Vegh:2013sk, Cai:2014znn}, and this allows people to study the holographic dual on the boundary\cite{Vegh:2013sk, Davison:2013jba, Blake:2013bqa}. In four dimensions, for certain such solutions with specific type of reference metrics, we derive the counterterms constructed by the invariants on the boundary. These counterterms cancel the divergences in the action and the boundary stress energy tensor. With these counterterms at hand, we can compute various quantities such as the thermodynamic potentials and the renormalized boundary stress energy  tensor and discuss holographic renormalization further in depth. Computation of various correlation functions can also be done.

Our method is basically the one reviewed in \cite{Skenderis:2002wp}. The difference is that we include odd powers of the radial coordinate in the expansions of the geometric quantities due to the existence of the massive gravity terms. Our analysis is in four dimensions, and to generalize the method in more generic dimensions one should consider the logarithmic terms to address the conformal anomalies. Meanwhile, in higher dimensions there are more massive terms and the case may be more complicated. If we consider more general forms of ${{\cal K}^\mu}_\nu$, probably we need to deal with different divergences. The method we used relies on some constraints such as (\ref{degenerating K}) and the regular expansions (\ref{expan of ma terms}) and so on. Constructing counterterms for more general cases may be a direction of future work. Moreover, apart from the dRGT theory there are other theories including momentum dissipation, such as the theories studied in the paper \cite{Baggioli:2014roa}. It is also interesting to study holographic renormalization in these theories.

By looking into the behavior of the boundary stress energy tensor, we find that the Ward identity still holds. However actually there is momentum dissipation on the boundary field theory and the translation symmetry is broken according to the literature\cite{Vegh:2013sk, Davison:2013jba, Blake:2013bqa}. There seems to be some difference but there is no contradiction. We have to emphasize that in their analysis the translation symmetry is violated in the context of perturbations. There may be a possibility that this violation does not show explicitly on the background level. This is not very strange since for this massive gravity model the diffeomorphism invariance in the bulk, dual to the translational symmetry on the boundary, is not manifestly broken. It is interesting to study the linear response of boundary fluids
by our renormalized stress energy tensor. Probably the dissipation of momentum will appear in this kind of discussion. This will be investigated in the near future.

\section*{ACKNOWLEDGMENTS}

This work was supported in part by the National Natural Science Foundation of China with grants
No.11205148 and No.11235010. This work was also Supported by the Open Project Program of State Key Laboratory of Theoretical Physics, Institute of Theoretical Physics, Chinese Academy of Sciences, China (No.Y5KF161CJ1).

\appendix
\section{The solutions of the equations of motion}\label{appeom}
Here we state the solutions of the equations of motion  in massive graivty order by order. Solving eq.(\ref{evolution}) to the leading order gives
\begin{eqnarray}
g_{(1)ij} &=& -\frac{1}{2}m^2 c_1\xi_{(0)ij}\label{g{(1)}}\,, \\
\mathrm{Tr}g_{(1)}&=&- \frac{1}{2} m^2 c_1 \mu_{(0)}\label{traceg{(1)}}\,,
\end{eqnarray}
and next to leading order we have
\begin{eqnarray}
g_{(2)ij} &=&\, \frac{3}{16}m^2 c_1 \mu_{(1)} g_{(0)ij} + \frac{1}{16}m^4 c_1^2 \mu_{(0)}\xi_{(0)ij} - \frac{1}{64}m^4c_1^2\mu_{(0)}^2 g_{(0)ij} \nonumber\\
& & \qquad   - \frac{1}{2}m^2c_1 \left( \xi_{(1)ij} - \frac{1}{4} \mathrm{Tr}\xi_{(1)}g_{(0)ij}\right) -\frac{1}{8} m^4 c_1^2 \xi_{(0)ik} \xi^k_{(0)j} \nonumber\\
& & + m^2 c_2 \left(X_{(0)ij} - \frac{1}{4}g_{(0)ij}\mathrm{Tr}X_{(0)}\right) - R_{(0)ij} + \frac{1}{4}R_{(0)}g_{(0)ij}\,,\\
\nonumber\\
\mathrm{Tr}g_{(2)} & =&\, \frac{1}{64}m^4c_1^2\mu_{(0)}^2 + \frac{1}{16}m^2 c_1 \mu_{(1)} - \frac{1}{8}m^2 c_1\mathrm{Tr}\xi_{(1)} + \frac{m^2 c_2}{4}\mathrm{Tr}X_{(0)}- \frac{1}{4}R_{(0)}\,.
\end{eqnarray}
Here $$X_{(0)ij}\equiv - \mu_{(0)} \xi_{(0)ij} + \xi_{(0)ik} \xi^k_{(0)j}\, ,$$ and all the indices are raised by $g_{(0)}^{ij}$. The $g_{(0)}^{-1}$s in the traces are omitted.
Moreover, we have these relations from (\ref{trace}) and the trace of (\ref{evolution})
\begin{equation}
\mathrm{Tr}\left(g_{(1)}^2\right) = m^2 c_1 \mu_{(1)}\label{order to order1}\,,
\end{equation}
\begin{equation}
-\frac{3}{2}\mathrm{Tr}\left(g_{(0)}^{-1}g_{(3)}\right) + \frac{1}{2}\mathrm{Tr}\left(g_{(1)}g_{(2)}\right) = -\frac{1}{4}m^2 c_1 \mu_{(2)}\,,
\end{equation}
\begin{eqnarray}
 && \mathrm{Tr}g_{(1)}\mathrm{Tr}g_{(2)} - \frac{1}{2}\mathrm{Tr}g_{(1)}\mathrm{Tr}\left( g_{(1)}^2 \right) + 5 \mathrm{Tr}\left( g_{(1)}g_{(2)} \right) - \frac{3}{2}\mathrm{Tr}\left( g_{(1)}^3 \right) - 6\mathrm{Tr}g_{(3)} \nonumber\\
 &&- \left(-g^{ki}_{(0)}g_{(1)kl}g^{lj}_{(0)}R_{(0)ij} + \mathrm{Tr}R_{(1)} \right) = m^2 c_1 \mu_{(2)} + m^2 c_2 \lambda_{(1)} \, .
\end{eqnarray}

\section{The coefficients of the divergences}
\label{appcoeff}

The coefficients of the divergent terms are obtained as follows by using the relations in appendix \ref{appeom}.
\begin{eqnarray}
a_{(0)} &=&\, 4\sqrt{-g_{(0)}}\,,
 \\
 \nonumber\\
a_{(1)} &=&\, 3\left.\frac{\partial \sqrt{-g}}{\partial u}\right|_{u=0} - \sqrt{-g_{(0)}} \mathrm{Tr}g_{(1)} - \sqrt{-g_{(0)}} \frac{1}{4} m^2 c_1 \mu_{(0)} \nonumber\\
&  =&\, - \frac{1}{2} \sqrt{-g_{(0)}} m^2 c_1 \mu_{(0)}\,, \\
\nonumber\\
a_{(2)} &=&\, -2\sqrt{-g_{(0)}} \mathrm{Tr}g_{(2)} - \sqrt{-g_{(0)}} \mathrm{Tr}\left(g^{-1}_{(1)}g_{(1)}\right) - \left.\frac{\partial \sqrt{-g}}{\partial u}\right|_{u=0} \mathrm{Tr}g_{(1)} \nonumber \\
&&  - \frac{m^2 c_1}{2} \sqrt{-g_{(0)}} \mu_{(1)} - \frac{m^2 c_1}{2} \left.\frac{\partial\sqrt{-g}}{\partial u}\right|_{u=0} \mu_{(0)}  \nonumber\\
 \nonumber\\
&=&\, -2\sqrt{-g_{(0)}} \mathrm{Tr}g_{(2)} + \frac{m^2 c_1}{2} \sqrt{-g_{(0)}} \mu_{(1)}\,,
\\
\end{eqnarray}
The coefficient $a_{(3)}$ can be proven to be vanishing,
\begin{eqnarray}
a_{(3)}&=&\, \left.\frac{\partial^3 \sqrt{-g}}{\partial u^3}\right|_{u=0} + \frac{m^2 c_1}{2} \left(\frac{\mu_{(0)}}{2}\left.\frac{\partial^2 \sqrt{-g}}{\partial u^2}\right|_{u=0} + \mu_{(1)}\left.\frac{\partial \sqrt{-g}}{\partial u}\right|_{u=0} + \mu_{(2)}\sqrt{-g_{(0)}}\right) \nonumber\\
\nonumber\\
&=&\, \sqrt{-g_{(0)}}\left( \frac{1}{8}m^4 c_1^2 \mu_{(0)}\mu_{(1)} + \mathrm{Tr}\left(g_{(1)}^3\right) - 2\mathrm{Tr}\left(g_{(1)}g_{(2)}\right) -\frac{1}{2}m^2 c_1 \mu_{(0)}\mathrm{Tr}g_{(2)} + m^2 c_1 \mu_{(2)}\right)\nonumber\\
& \propto &\, \frac{1}{8}m^4 c_1^2 \mu_{(0)}\mu_{(1)} + \frac{1}{4}\left( -\frac{1}{8} \right) m^6 c_1^3 \xi^i_{(0)k} \xi^k_{(0)l}\xi^l_{(0)i} + \frac{1}{64}m^4 c_1^2 \mu_{(0)}\mu_{(1)} \nonumber\\
&& + \frac{1}{64}m^6 c_1^3 \mu_{(0)} \xi^i_{(0)j} \xi^{j}_{(0)i}  - \frac{1}{32}m^6 c_1^3 \xi^i_{(0)k} \xi^k_{(0)l}\xi^l_{(0)i} \nonumber\\
&&\, - \frac{1}{8}m^4 c_1^2 \left(\xi_{(1)ij} \xi^{ji}_{(0)} - \frac{1}{4}\mathrm{Tr}\xi_{(1)}\mu_{(0)}\right) + \frac{m^4 c_1 c_2}{4}\left( X^i_{(0)j} \xi^{j}_{(0)i} - \frac{1}{4}\mu_{(0)}\mathrm{Tr}X_{(0)}\right)\nonumber\\
&&\, - \frac{3}{4}m^2 c_1 \mu_{(0)}\left( \frac{1}{8}m^2 c_1\mu_{(1)} + \frac{1}{16}m^2 c_1\mu_{(1)} - \frac{1}{8}m^2 c_1\mathrm{Tr}\xi_{(1)} + \frac{1}{4}m^2 c_2 \mathrm{Tr}X_{(0)} \right) \nonumber\\
&&\, - \frac{1}{4}\left(-\frac{1}{2} m^2 c_1 \mu_{(0)}\right) m^2 c_1 \mu_{(1)} - \frac{1}{2}m^2 c_2 \lambda_{(1)} - \frac{1}{2} \mathrm{Tr}\left(g_{(1)}^{-1}R_{(0)}\right) - \frac{1}{2}R_{(0)}\mathrm{Tr}g_{(1)} \nonumber\\
&&\,- \frac{1}{2}\left(\mathrm{Tr}\left(g_{(1)}^{-1}R_{(0)}\right) + \mathrm{Tr} R_{(1)}\right)\nonumber\\
&=&0 \,.
\end{eqnarray}
It is vanishing partly because of the relations (\ref{degenerating K}) and (\ref{trace squared}). One can use these relations to find a relation between $\mu_{(0)}$ and $\mu_{(1)}$:
\begin{equation}
8 m^2c_1\mu_{(1)} = m^4 c_1^2 \mu_{(0)}^2\,,
\end{equation}
and the relations between the various contractions of $\xi_{(l)ij}$s, and $\mu_{(0)}$.
Then one can simplify the expression a lot. By expressing $R_{(1)}$ in terms of $g_{(0)}$ and $g_{(2)}$ ( this trick is given in detail in the appendix of \cite{Kanitscheider:2008kd}) and considering (\ref{metric divergence eq})  one can find that $\mathrm{Tr} R_{(1)}=0$. Then there are only the $R_{(0)ij}$ terms left. We can impose the near boundary condition for an asymptotically AdS space (see, for instance \cite{Hollands:2005wt}) that the boundary metric $g_{ij}$ is an Einstein static universe:
\begin{equation}
g_{ij} = -dt^2 + r^2 \Sigma_{ij}dx^i dx^j\,,\qquad i,j= 2, 3\,,
\end{equation}
where $\Sigma_{ij}$ is the metric of a maximally symmetric $2$-surface.
Then $R_{(0)ij}$ can be computed explicitly. We can choose $\sigma_{ij}$ in (\ref{degenerating K}) near the boundary to be $\Sigma_{ij}$ (very naturally), and get that the $R_{(0)ij}$ terms in $a_{(3)}$ cancel.

Finally, the divergent part of the action consists of the following parts:
\begin{eqnarray}
a_{(0)} \epsilon^{-3} & \approx &\, 4\sqrt{-\gamma} \left(1 + \epsilon\frac{1}{4}m^2 c_1 \mu_{(0)} \right. \nonumber\\
&& \qquad \left.\, + \epsilon^2 \left(\frac{1}{32}m^4 c_1^2 \mu_{(0)}^2 + \frac{1}{4}m^2 c_1 \mu_{(1)} - \frac{1}{2}\mathrm{Tr}g_{(2)}\right)\right)\, , \\
\nonumber\\
a_{(1)} \epsilon^{-2} & \approx &\,  -\sqrt{-\gamma} \, \epsilon\left(1 + \epsilon\frac{1}{4}m^2 c_1 \mu_{(0)}\right) \frac{1}{2} m^2 c_1 \mu_{(0)}\, , \\
\nonumber\\
a_{(2)} \epsilon^{-1} & \approx &\, \sqrt{-\gamma} \, \epsilon^2 \left(-2 \mathrm{Tr}g_{(2)} + \frac{m^2 c_1}{2} \mu_{(1)}\right)\, ,
\end{eqnarray}
all of which are up to $\mathcal{O}(1)$\,. The sum of these divergences is
\begin{eqnarray}
\label{structure}
S_{\mathrm{div}}&=&\frac{1}{2\kappa^2}\int_{\partial M} d^3 x \sqrt{-\gamma} \left[4 + \epsilon \frac{1}{2}m^2 c_1 \mu_{(0)} + \epsilon^2 \frac{1}{2}m^2 c_1 \mu_{(1)} + \epsilon^2 \left(m^2 c_1 \mu_{(1)} - 4\mathrm{Tr}g_{(2)}\right) \right] \nonumber\\
&=& \frac{1}{2\kappa^2}\int_{\partial M} d^3 x\sqrt{-\gamma}\left[4 + \epsilon \frac{1}{2}m^2 c_1 \mu_{(0)} + \epsilon^2 \frac{1}{2}m^2 c_1 \mu_{(1)} \right. \nonumber\\
&& \left. \qquad + \epsilon^2 \left( -\frac{1}{16}m^4 c_1^2 \mu_{(0)}^2 + \frac{1}{16} m^4 c_1^2 \xi_{(0)ij}\xi_{(0)}^{ji} - m^2 c_2 \mathrm{Tr}X_{(0)}+R_{(0)}[g] \right) \right] \nonumber\\
&\approx &\frac{1}{2\kappa^2}\int_{\partial M} d^3 x\sqrt{-\gamma}\left[4 + R[\gamma] + \frac{1}{2}m^2 c_1 {\cal U}_1 + \left(m^2 c_2 - \frac{1}{16}m^4 c_1^2\right) {\cal U}_2  \right]\,.
\end{eqnarray}
So the counterterm with $\ell$ recovered is given by eq.(\ref{countertermaction}). Here we have used eq.(\ref{g{(1)}}), eq.(\ref{traceg{(1)}}), eq.(\ref{order to order1}) and (\ref{trace squared}).

\section{The divergence of $T_{ij}$}
\label{app div T}
We show here that the divergence of $T_{ij}$ is zero. The first term in the second line of (\ref{divergence of energy tensor}) is proportional to
\begin{align}
&\quad\, D^\mu\left( {\cal U}_1 \gamma_{\mu\nu} - {\cal K}_{\mu\nu} \right) \nonumber \\
&=\nabla^\mu\left( {\cal U}_1 \gamma_{\rho\sigma} - {\cal K}_{\rho\sigma} \right) {\gamma^\rho}_\mu {\gamma^\sigma}_\nu \nonumber\\
&=\nabla^\mu\left( {\cal U}_1 {\cal G}_{\rho\sigma} - {\cal U}_1 n_\rho n_\sigma - {\cal K}_{\rho\sigma} \right){\gamma^\rho}_\mu {\gamma^\sigma}_\nu \nonumber\\
&=\nabla^\mu\left( {\cal U}_1 {\cal G}_{\rho\sigma} - {\cal K}_{\rho\sigma} \right) {\gamma^\rho}_\mu {\gamma^\sigma}_\nu \nonumber\\
&=\nabla^\mu\left( {\cal U}_1 {\cal G}_{\mu\sigma} - {\cal K}_{\mu\sigma} \right)  {\gamma^\sigma}_\nu - \nabla^\mu\left( {\cal U}_1 {\cal G}_{\rho\sigma} - {\cal K}_{\rho\sigma} \right)n^\rho n_\mu {\gamma^\sigma}_\nu \nonumber\\
&= 0 -\nabla^\mu\left( {\cal U}_1 {\cal G}_{\rho\sigma} - {\cal K}_{\rho\sigma} \right)  n^\rho n_\mu {\gamma^\sigma}_\nu \nonumber\\
&= 0 + \nabla_\mu {\cal K}_{\rho\sigma} n^\rho n_\mu {\gamma^\sigma}_\nu\nonumber\\
&= - {\cal K}_{\rho\nu} n^\mu \nabla_\mu n^\rho = 0\,,
\end{align}
where $\nabla$ is the covariant derivative compatible with the spacetime metric $\cal G$.
The last line vanishes because the vector $-(1/u) n^\mu$ generates geodesics in the conformal spacetime with the unphysical metric $u^2\mathcal{G}_{\mu\nu}$ so after the conformal transformation to the physical metric $\cal G$ the quantity $n^\mu \nabla_\mu n^\rho$ is still proportional to $n^\rho$. We have also used the property that each single term with coefficient $c_l$ in the RHS of (\ref{eom}) is conserved so
\begin{equation}
\nabla^\mu\left( {\cal U}_1 {\cal G}_{\mu\sigma} - {\cal K}_{\mu\sigma}\right)=0\,.
\end{equation}
As to the second term in (\ref{divergence of energy tensor}) the proof is almost the same. Therefore $T_{ij}$ is divergenceless.

\end{document}